\documentclass[twoside,showpacs,showkeys]{elsarticle} 

\usepackage[utf8]{inputenc}
\usepackage{amsmath,amsfonts,amssymb}
\usepackage{graphicx}
\usepackage[english]{babel}
\usepackage[svgnames]{xcolor}

\usepackage{hyperref}
\hypersetup{colorlinks=true,urlcolor=Green,linkcolor=blue}

\definecolor{color1}{RGB}{0,0,100} 

\newlength{\tocsep} 
\def\vv{\textbf}

\begin{document}


\title{Vulnerability of state-interdependent networks under malware spreading}

\author[upco]{Rafael Vida}
\ead{rvida@upcomillas.es}
\author[upm]{Javier Galeano}
\ead{javier.galeano@upm.es}
\author[uam]{Sara Cuenda\corref{cor1}}
\ead{sara.cuenda@uam.es}

\cortext[cor1]{Corresponding author}

\address[upco]{Dept. de Sistemas Inform\'aticos, Escuela T\'ecnica Superior de Ingenier\'{\i}a (ICAI), C/ Alberto Aguilera 25, 28015 Madrid, Spain and Grupo Interdisciplinar de Sistemas Complejos (GISC)}
\address[upm]{Dept. Ciencia y Tecnolog\'{\i}a Aplicadas a la I.T. Agr\'{\i}cola,
E.U.I.T. Agr\'{\i}cola, Universidad Polit\'ecnica de Madrid, 28040 Madrid, Spain and Complex Systems Group (GSC)}
\address[uam]{Dept. Econom\'{\i}a Cuantitativa, Universidad Aut\'onoma de Madrid, C/ Francisco Tom\'as y Valiente 5, 28049 Cantoblanco (Madrid), Spain and Grupo Interdisciplinar de Sistemas Complejos (GISC)}

\begin{abstract}
Computer viruses are evolving by developing spreading mechanisms based on the use of multiple vectors of propagation. The use of the social network as an extra vector of attack to penetrate the security measures in IP networks is improving the effectiveness of malware, and have therefore been used by the most aggressive viruses, like Conficker and Stuxnet. In this work we use interdependent networks to model the propagation of these kind of viruses. In particular, we study the propagation of a SIS model on interdependent networks where the state of each node is layer-independent and the dynamics in each network follows either a contact process or a reactive process, with  different propagation rates. We apply this study to the case of existing multilayer networks, namely a Spanish scientific community of Statistical Physics, formed by a social network of scientific collaborations and a physical network of connected computers in each institution. We show that the interplay between layers increases dramatically the infectivity of viruses in the long term and their robustness against immunization. 
\end{abstract}

\begin{keyword}
networks \sep multiplex networks \sep interdependent networks \sep Markov processes \sep contagion spreading \sep percolation

\PACS 02.50.Ga \sep 89.75.-k \sep 89.20.-a
\end{keyword}



\maketitle 


\section{Introduction}

In response to a request from the UK Ministry of Defense, Anderson and coworkers estimated the global cost of malware at US\$370 millions in the 2010 year \cite{2012:anderson}. In this report, they explain that some of the reasons of the inefficiency of war against cybercrime is that malware is global and have strong externalities. In this sense, during the last years computer viruses have developed complex spreading mechanisms that allow them to propagate using several mechanisms. There are noted examples, like Conficker \cite{porras2009conficker} or Stuxnet \cite{falliere2011w32}, which had an enormous impact in the Internet network and use these new methods of spreading. For these kind of viruses the propagation is easy and quick within a Local Area Network (LAN). However, effective security measures \cite{antoine2003router} limit the propagation of these viruses to other LANs. To overcome this limit, these viruses also make use of other secondary vector of propagation such as the social relations between humans. Due to the complexity of the virus propagation and infection, re-infection is quite common after virus removal, so it is technically complicated to clean a whole LAN quickly enough to stop the re-infection. 

Contagion and epidemic spreading have been widely studied in the scientific literature, usually considering only one network \cite{2001:pastora,2001:pastorb,2001:may,2002:moreno,2010:gomez,2011:gleeson,2009:vanmieghem,klemm2012} and, more recently, using several interconnected layers of networks and multiplexes \cite{2010:buldyrev,2012:son,2012:boguna,2013:granell,2013:yamir,2014:guha} and several infectious agents \cite{2010:funk,2011:marceau}. However, none of these formalisms suits the case we are dealing with, namely a single disease which spreads over a set of agents which are interconnected through several networks, each with a different propagation regime, but in which the state of each agent in every network must be the same. This scenario is quite common in disease propagation. Opinions may circulate around society, but each network of social ties (family, close friends, work-mates, followees,\ldots) affects differently our opinion depending on the contact rate and our trust. Similarly, human diseases such as flu or venereal diseases propagate with rates of infection that clearly depend on social relationships. In the case of malware spreading computers are usually connected within a local network and also through a social network of contacts that involve receiving corporate, private and spam e-mails or plugging foreign pen drives in the computers. 

In this paper we develop a new formalism that applies to the study of the epidemic spreading in these kind of systems. These can be understood as a special subset of {\em interdependent networks} \cite{2014:boccaletti} with no explicit links joining the networks but where the state of any node must be the same in every layer. 
\footnote{Other references to previous work and nomenclature can be found in \cite{2013:kivela}.}
We will hereafter call them state-interdependent networks (SINs).
Our case study is the propagation of a SIS epidemic model in SINs. We show that the disease dynamics can be described in terms of a single contagion matrix that subsumes the contagion processes of all layers. This matrix can be used to calculate any node or link-dependent magnitude concerning the epidemic spreading such as the centrality of nodes or links, the community structure of the disease, {\it etc}. Finally, based on our formalism, we show the effect of some immunization strategies to slow down or control the epidemic dynamics.
An important part of our analysis includes the study of an actual SIN, a Spanish scientific community of Statistical Physics which is connected through the social network of scientific collaborations and the physical network of the university LANs, and simulate the spreading process of a SIS disease.

\section{The model} 

\begin{figure}
	\centering
	\includegraphics[width=0.5\columnwidth]{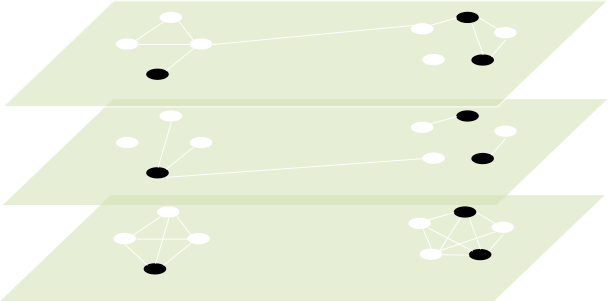}
	\caption{\label{fig:multi-layer}
	State interdependent networks (SINs), where the state of each node in every layer is the same but the interactions within each layer may differ. Infected nodes are represented in black and susceptible nodes in white.
	}
\end{figure}

Let us consider $M$ layers of networks formed each one by $N$ nodes. The usual adjacency matrix is replaced by a set of matrices, $A^{(\alpha)}=\left(A^{(\alpha)}_{ij}\right)$ with $\alpha=1,\ldots,M$, that specifies the links between nodes in each layer $\alpha$. Note that, in these SINs, the state of nodes with the same label must be the same, and the change in the state of one node in one layer changes automatically his state in all other layers (see figure \ref{fig:multi-layer}).

In these SINs we will study a SIS epidemic spreading in which the contagion in every layer $\alpha$ may propagate differently. We will assume that the epidemic spreading in each layer may follow a contact process, a reactive process or something in between \cite{2010:gomez}. To this end we define the contagion matrix $C^{(\alpha)}=( C^{(\alpha)}_{ij})$ in layer $\alpha$ as 
\begin{equation}\label{eq:rij}
	C^{(\alpha)}_{ij}=\beta^{(\alpha)}_i\left(1-\left(1-\frac{w^{(\alpha)}_{ij}}{w^{(\alpha)}_{i}} \right)^{\lambda^{(\alpha)}_i}\right),
\end{equation}
where $w^{(\alpha)}_{ij}$ stands for the weight of the link between node $i$ and $j$, $w^{(\alpha)}_{i}=\sum_j w^{(\alpha)}_{ij}$ is the total strength \cite{2005:barthelemy} of node $i$, $\beta^{(\alpha)}_i$ is a constant between $0$ and $1$, and $\lambda^{(\alpha)}_i$ is the parameter that defines the contagion process for node $i$, which varies from a reactive process for the limit $\lambda^{(\alpha)}_i\rightarrow \infty$ to a contact process for $\lambda^{(\alpha)}_i=1$. 

The system state is described by the vector state $\vv{x}=\{x_1,\ldots,x_N\}$, with $x_i=0$ when node $i$ is susceptible and $x_i=1$ when is infected. The transition rate for node $i$ from infected to susceptible is
	\begin{align}
		q^-_i(\vv{x})&=\mu x_i \quad \label{eq:trans-}
	\end{align}
where $\mu$ is the recovery rate, which we assume layer-independent since the same healing mechanisms are available to all nodes (this assumption can, however, be easily relaxed). On the other hand, the transition rate from susceptible to infected is 
	\begin{align}
		q^+_i(\vv{x})&=\sigma(1-x_i)\left[1- \prod_{\alpha=1}^M\prod_{j=1}^N
		(1- C^{(\alpha)}_{ji}x_j)\right], \label{eq:trans+}
	\end{align}	
where $\sigma$ has units of [time]$^{-1}$.
In expression \eqref{eq:trans+}, all the possible contagions from the infected neighbors of node $i$ in all layers have been considered. With these transition rates we can express the epidemic model as a Markov process in continuous time following the master equation
\begin{equation}\label{eq:markov}
	\begin{split}
	\frac{\partial P(\vv{x},t)}{\partial t} =&
	\sum_{i=1}^N\left\{\left[
	q_i^{+}(f_i(\vv{x}))+q_i^{-}(f_i(\vv{x})) \right]P(f_i(\vv{x}),t)\right.\\
	&\quad\quad\left.-\left[q_i^{+}(\vv{x})+q_i^{-}(\vv{x})	\right]P(\vv{x},t)
	\right\},
\end{split}
\end{equation}
where $f_i(x_1,\ldots,x_i,\ldots,x_N)=(x_1,\ldots,1-x_i,\ldots,x_N)$ is the flip operator of the $i$-th component, that changes the state of node $i$ from susceptible to infected and vice versa.

Expressions \eqref{eq:rij} and \eqref{eq:trans+} account for independent contagion processes that take place in every layer concurrently. This assumption is valid in the context of modern malware spreading, where no competition between layers exists. Each layer contributes equally to the contagious process and no dilution between layers is accounted for explicitly.

Notice that the definition of expression \eqref{eq:trans+} is such that $C^{(\alpha)}_{ji}$ must be a probability, ranged between $0$ and $1$, and not a probability rate, in order for $q_i^{+}(\vv{x})$ to be well defined. From now on, we simply assume that $\sigma=1$, which amounts to choosing an appropriate time scale in the master euqation \eqref{eq:markov}.
As a result, $\mu$ represents the ratio between the transtion rates from infected to susceptible and viceversa, without loss of generality.

\subsection{Effective contagion matrix} 
Since the variables $x_{j}$ are binary with $x_j\in\{0,1\}$, it holds that $1-C^{(\alpha)}_{ji} x_j=(1- C^{(\alpha)}_{ji})^{x_j}$, which yields
$\prod_\alpha\prod_j (1-C^{(\alpha)}_{ji} x_j)=
\prod_j\prod_\alpha (1- C^{(\alpha)}_{ji})^{x_j}$. This allows the definition of the {\em effective contagion matrix}, 
\begin{equation}\label{eq:Reff}
	\bar{C}_{ij}\equiv 1-\prod_{\alpha=1}^M(1-C^{(\alpha)}_{ij}),
\end{equation}
and therefore the transition rate $q^+_i(\vv{x})$ is expressed as
\begin{equation}\label{eq:trans2}
	q^+_i(\vv{x})=(1-x_i)\left[1- \prod_{j=1}^N 	(1-\bar{C}_{ji}x_j)\right].
\end{equation}
Notice that expression \eqref{eq:trans2} has no explicit layer dependence, and thus $\bar{C}$ can be interpreted as the contagion matrix that would render the same node dynamics in a single network as the one defined in the SINs.

Since the dynamics on SINs is ruled by the effective contagion matrix \eqref{eq:Reff}, all the system properties must be obtained from $\bar{C}$. For instance, its maximum eigenvalue $\bar{\Lambda}_{\max}$ is related to the onset of the disease \cite{2009:vanmieghem,2010:gomez}. The left eigenvector associated to $\bar{\Lambda}_{\max}$, $\vv{p}=(p_1,\ldots,p_N)$, approximates the expected probabilities for node $i$ to be infected in the limit of (independent) small probability \cite{2010:gomez} and is also related to the dynamical influence of each node to the rest of the network in the contagion process \cite{klemm2012}. The maximum eigenvalue is bounded in every layer $\alpha$ by the expression
\begin{equation}\label{eq:eig}
	\Lambda^{(\alpha)}\le
    \frac{\vv{p}^{(\alpha)} \bar{C}\vv{q}^{(\alpha)}}{\vv{p}^{(\alpha)}\vv{q}^{(\alpha)}}\le
	\bar{\Lambda}_{\max}\le 
	\frac{\vv{p} \left(\sum_{\alpha} C^{(\alpha)} \right)\vv{q}}{\vv{p}\vv{q}}\le
	\sum_{\alpha} \Lambda^{(\alpha)}
\end{equation}
where $\vv{q}$ is the dual right eigenvector of $\vv{p}$ and $\Lambda^{(\alpha)}$ is the maximum eigenvalue of $C^{(\alpha)}$ with left eigenvector $\vv{p}^{(\alpha)}$ and dual right eigenvector $\vv{q}^{(\alpha)}$.

Expression \eqref{eq:eig} renders $\max_{\alpha}(\Lambda^{(\alpha)})\le \bar{\Lambda}_{\max}\le \sum_{\alpha} \Lambda^{(\alpha)}$. These bounds impose faster dynamics, more infectious results and lower epidemic onsets for a virus propagating in the SINs than in any isolated layer.
The bounds on $\bar\Lambda_{\max}$ are a consequence of the model we are using, which assumes that the spreading processes are independent within each layer, and that there is no dilution between layers when coupling the networks by the nodes. Under different assumptions these bounds may not hold.
  
Expression \eqref{eq:Reff} applied to the case of two layers renders 
\begin{equation}\label{eq:Reff2}
	\bar{C}_{ij}=C^{(1)}_{ij}+C^{(2)}_{ij}-C^{(1)}_{ij}C^{(2)}_{ij}.
\end{equation}
Since $0\le C_{ij}^{(\alpha)} \le 1 $, it follows that $
	\max\left(C^{(1)}_{ij},C^{(2)}_{ij}\right)\le 
	\bar{C}_{ij}\le \min\left(1,C^{(1)}_{ij}+C^{(2)}_{ij}\right),$
which can be easily extended to the general SIN case by induction,
\begin{equation*}
  \max_{\alpha}\left(C^{(\alpha)}_{ij}\right)\le 
	\bar{C}_{ij}\le \min\left(1,\sum_\alpha C^{(\alpha)}_{ij}\right)
\end{equation*}

\begin{figure}
	\centering
	\includegraphics[width=0.5\columnwidth]{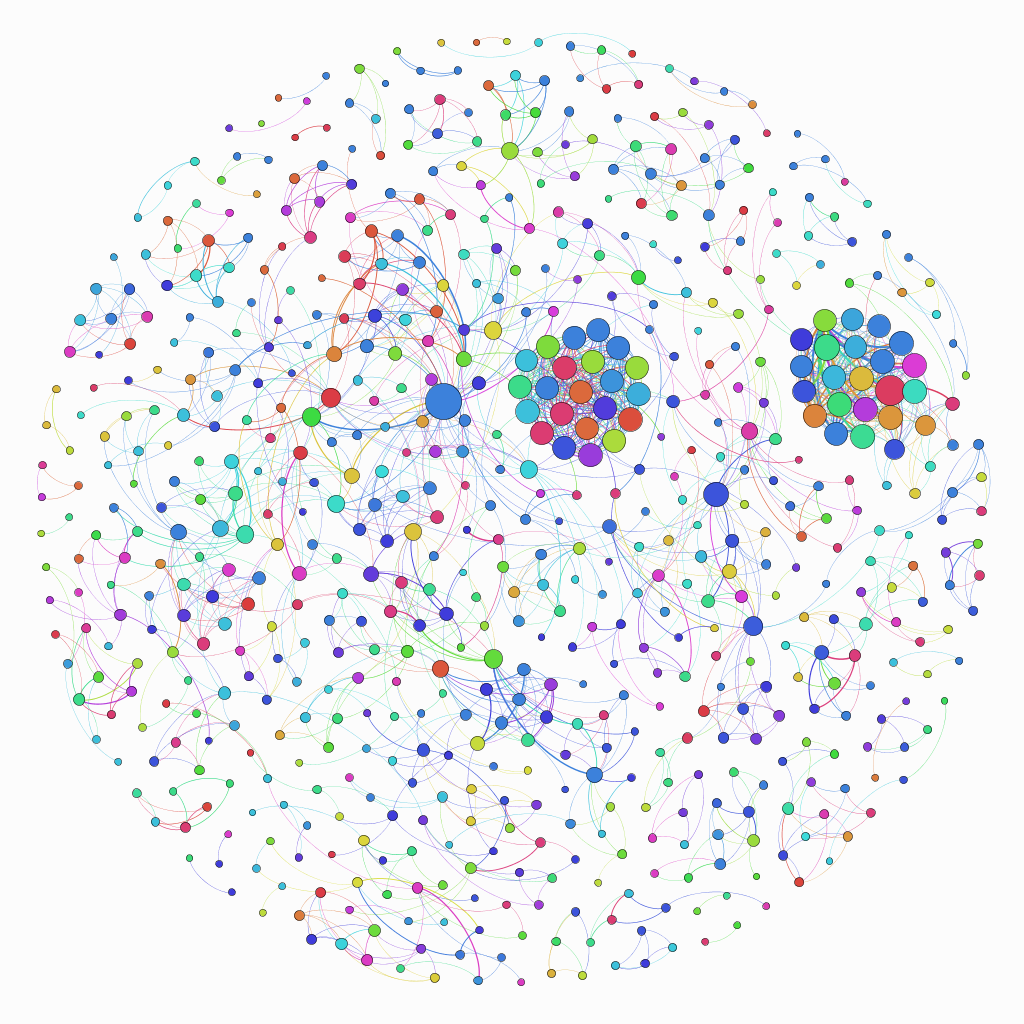}
	\caption{\label{fig:uno}
	(Color online) Social network of the Spanish Statistical Physics scientific community (FISES). Links represent collaborations between authors, the size of each node the number of links and the color indicates the affiliation. Therefore, links joining nodes of different colors show collaborations between research institutions.
	}
\end{figure}

There are two special cases that we would like to discuss in order to give some insight on the interplay of just two layers. In the case that both layers have exactly the same network topology (i.e., the same adjacency matrix $A$) and layer $1$ follows a reactive process with $C^{(1)}=A$, then for any $C^{(2)}=(C^{(2)}_{ij})$ satisfying $C^{(2)}_{ij}=b_{ij}A_{ij}$ we find from equation \eqref{eq:Reff2} that
$$
\bar{C}_{ij}=A_{ij}+b_{ij}A_{ij}-b_{ij}A_{ij}^2=A_{ij}(1+b_{ij}-b_{ij})=A_{ij},
$$
where we have used that $A_{ij}^2=A_{ij}$. Therefore, since $\bar{C}=C^{(1)}$ in this particular case, the effect of the second layer in the contagion matrix (and, therefore, in the epidemics dynamics) vanishes.
This example, in which layer $2$ has no influence whatsoever in the dynamics of the SINs, shows that the effects on the dynamics of this system are not as simple as the addition of each layer effects. The second case is the only one in which the contagion matrix of the system is the sum of the contagion matrices of each separated network: two layers in which the intersection of the sets of edges of each layer is the empty set (i.e., there are no common links in both layers), and therefore $C^{(1)}_{ij}C^{(2)}_{ij}=0$ for every pair of nodes of the system.

\section{University LANs and networks of scientific collaborations} 
Firms are very cautious in sharing information that would make them seem vulnerable to rivals or potential shareholders and hackers. Thus public data are {\em rara avis} in this field of research. Instead of comparing model simulations to real data we can use the model to predict the hypothetical spreading of modern malware over real SINs that resemble the main features exploited by these kind of virus. Simulations of epidemic spreading in real networks have been developed in, for example, \cite{2006:colizza,2007:colizza,2010:kitsak,2013:chen}.

In this spirit, we consider the tandem formed by institutions' LANs and  scientific collaborations as a paradigmatic example of SINs. Usually, universities have one or more LANs, and each university connects its own LANs using IP switches or routers. The internal IP-nodes of each university are considered trusted, whereas the external IPs are considered possibly dangerous. Therefore, the connections with other universities or external LANs use secured links with firewalls as a way to implement the perimetral security controls, as well as IDS (intrusion detection system) or IPS (intrusion prevention system) to control the traffic coming to the internal LANs. With these techniques research institutions avoid most of the external attempts of malware infection.

This strategy of prevention by isolating small networks breaks down when we consider the social interactions of scholars and researchers. In particular, scientific collaborations involve a wide scope of social interactions, such as e-mails, virtual or face-to-face meetings, research visits, invited talks, or conference attendances, among others. Some of these interactions include connecting a foreign laptop to a local network (by wire or Wi-fi connections) or connecting a third party's pen drive to a computer. For instance, some authors relate the origin of the infection of Stuxnet with one SCADA (supervisory control and data acquisition) conference, as this SCADA systems were the primary target of the infection. It was attached to pen drives that also contained software that was distributed in the conference \cite{matrosov2010stuxnet}. 

\subsection{The FisEs community}
For these reasons, and since the information about collaborations and affiliations is publicly available, we have chosen the Spanish Statistical Physics (FisEs) research community. Eighteen periodic meetings since 1987 in a widespread of host universities have consolidated this community over the years \footnote{\url{http://www.fises.es}}. In this investigation we have used the contributions specified in the programs of the last two meetings to build up the network of scientific collaborations. In this network (the {\em social network}, in the following) two authors are linked if they have at least one common contribution to any of the two meetings. In the network of LANs (the {\em physical network}, from now on) two authors are linked if they are affiliated to the same university and department or to the same research center. Notice that the physical network is formed by disconnected cliques.

We obtained 345 contributions from 687 authors distributed in 105 affiliations, yielding 105 disconnected cliques in the physical network, the largest containing 39 nodes. With respect to the social network, there are 73 connected components, the largest formed by 188 nodes. One of the most striking features of the coupled network is that the number of connected components reduces drastically to 8, and the largest component has 657 nodes, almost the total number of the considered nodes \footnote{Data is available at \url{http://www.uam.es/sara.cuenda/research/fises_data.tgz}.}. Notice that, by coupling two sparsely connected networks we have obtained a network that reduces in one order of magnitude the number of connected components and with a largest component which is of the order of the total number of nodes. Figure \ref{fig:uno} shows the interconnections that the social network adds to the physical network (see caption for details). A remarkable result is that the combination of two highly disconnected networks renders an almost fully connected, highly clustered network. 
\begin{figure}
	\centering
	\includegraphics[width=0.5\columnwidth]{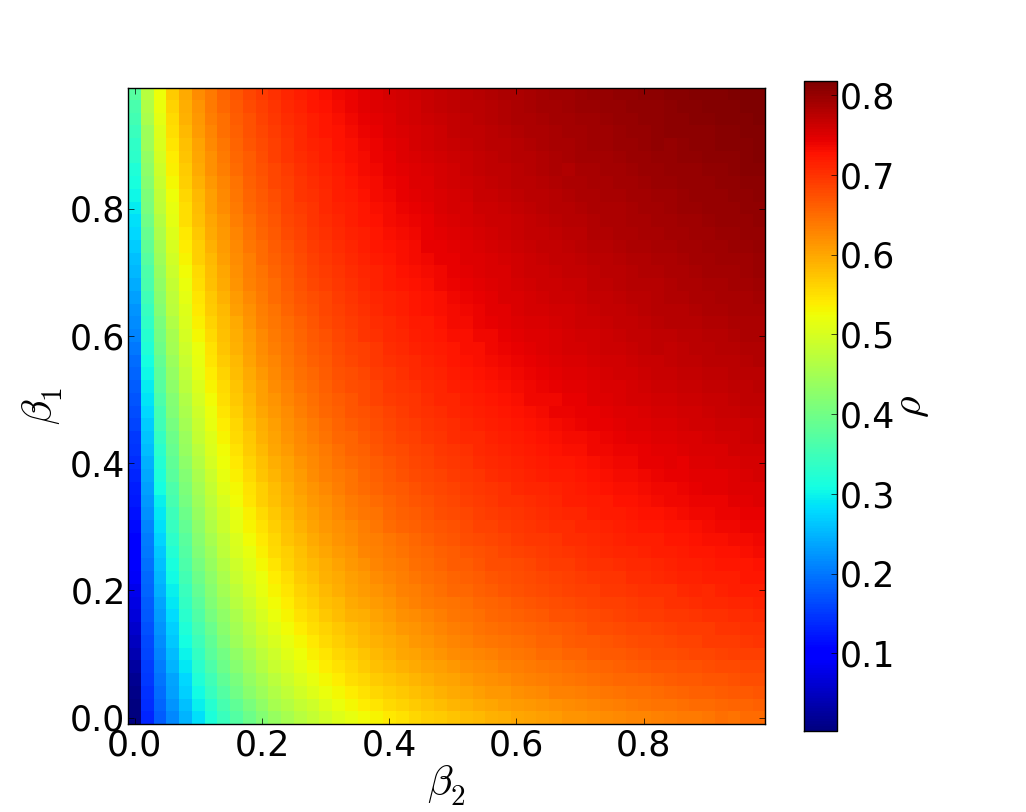}
	\caption{\label{fig:dos}
	(Color online) Density of infected nodes for large $t$ of the coupled network ($\rho$) in terms of the contagion rates of the social ($\beta_1$) and the physical ($\beta_2$) network, with $\mu=0.5$. See text for details. }

\end{figure}

\begin{figure}
	\centering
	\includegraphics[width=0.5\columnwidth]{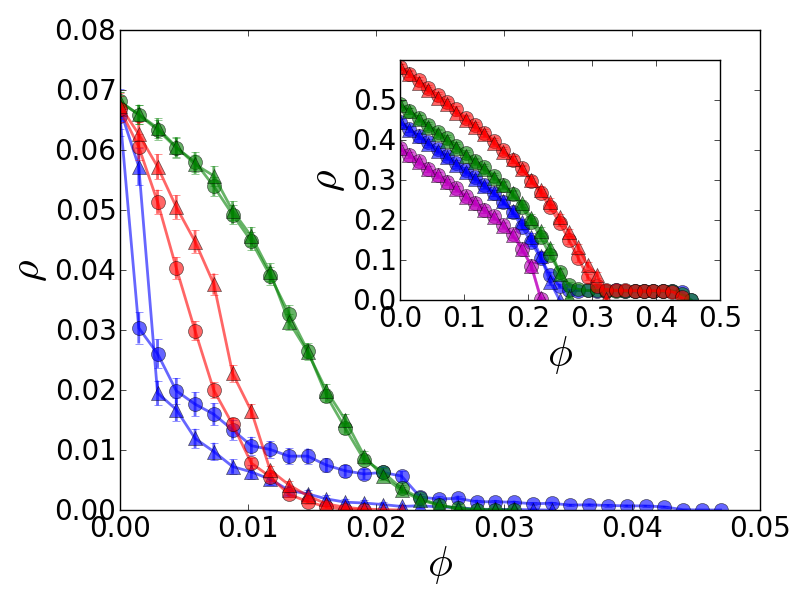}
	\caption{\label{fig:tres}
	Density of infected nodes $\rho$, vs.\ density of immunized nodes $\phi$. Triangles ($\triangle$) stand for an immunization strategy based on the strength of nodes in the contagion matrix $\bar{C}$, whereas circles ($\circ$) use the left eigenvector centrality of $\bar{C}$. The contagion rates are $\beta_1=1.0$, $\beta_2=0.0$ (blue); $\beta_1=0.0$, $\beta_2=0.027$ (green) and $\beta_1=0.18$, $\beta_2=0.02$ (red). Inset: The same, with contagion rates $\beta_2=0.2$ and $\beta_1=0.0$ (magenta); $\beta_1=0.1$ (blue); $\beta_1=0.2$ (green) and $\beta_1=0.5$ (red).}
\end{figure}

\subsection{Numerical results}

Since each layer may follow a different contagion process, the time scales of $q_i^+$ and $q_i^-$ defined in \eqref{eq:trans-} and \eqref{eq:trans+} may vary considerably. To account for this issue we numerically simulated trajectories $\vv{x}(t)$ associated to the embedded Markov chain of the continuous-time Markov process defined in \eqref{eq:markov} (see \cite{markov}), and averaged over 200 realizations for large $t$. All simulations start from the fully infected state to try to avoid more frequently the absorbing state of the system where all nodes are susceptible to the infection.


Figure \ref{fig:dos} shows the density of infected nodes for large $t$ in terms of the contagion rates of the social and the physical network. We have considered that the social meeting spreading mechanism can be modeled by a contact process since the infectivity of each infected node is divided among its neighbors. In the physical layer, however, the virus is constantly attacking all node's neighbors with the same intensity, regardless its connectivity, and therefore we have modeled this layer as a reactive process. 
The vulnerability of the whole system under this kind of malware becomes apparent in figure \ref{fig:tres}, where the density of infected nodes, $\rho$, vs.\ the density of immunized nodes, $\phi$, is shown for different values of the contagion rates $\beta_1$ and $\beta_2$ (see caption for parameter details). We have studied the effect on $\rho$ of the immunization of a fraction of nodes $\phi$ using two different strategies: by immunizing the ``strongest'' node (the one with higher strength $s_i=\sum_j \bar{C}_{ij}$) and by protecting the one with largest left eigenvector centrality in the largest connected component \cite{klemm2012}. The procedure is as follows: in each step, if $\ell$ is the node with the greatest value of strength or eigenvector centrality (depending on the strategy that we are using) \footnote{If several nodes have the greatest value, we choose one of them randomly.}, we immunize it by making $\bar{C}_{ij}=0$ for all $i=\ell$ or $j=\ell$; finally, we calculate $\rho$ for large $t$ and proceed with the next node.

The results for the two strategies are very similar, as can be seen in figure \ref{fig:tres}. In its main panel we compare the effect of immunization in the coupled and the uncoupled networks, choosing contagion rates such that all systems have the same density of infected nodes $\rho$ for $\phi=0$. Notice the differences in the choice of rates $\beta_1$ and $\beta_2$ in order to achieve this condition. 
The inset of figure \ref{fig:tres}, where several immunization processes have been simulated for a fixed value of $\beta_2$, shows that the physical layer, with high connectivity clusters and a reactive process, confers on the virus great spreading capacity, and the social layer enhances this robustness by adding links that let the virus spread to small institutions where otherwise the infection would have died out faster.

\section{Conclusions}
We have shown that the dynamics of a SIS contagion process in SINs where the state of nodes must be layer-independent is equivalent to the spreading in a mono-layer which is governed by the effective contagion matrix $\bar{C}$, which allows to treat the epidemic spreading as in a single network without introducing any approximation. We can therefore apply any of the previous works regarding SIS epidemics spreading on networks available in the literature \cite{2001:pastora,2001:pastorb,2001:may,2002:moreno,2010:gomez,2011:gleeson,2009:vanmieghem,klemm2012}.

We chose the pair formed by the universities LANs and the scientific collaborations as a paradigmatic example of the interplay between these two layers in the propagation of recent computer viruses. The construction of these networks must be understood as a way to obtain existing SINs which partially resemble the spreading mechanism of modern malware. This mechanism focuses on the multilayer feature of the system in order to connect small networks that otherwise would be isolated (both in the social and the physical network). In fact, we have not included other layers that would increase the connectivity among the researchers and add more nodes to interact with, increasing the infectivity of the disease. However, as we show in the numerical results, the two layers considered in our study are enough to dramatically increase the vulnerability of the system to infections. This result is in agreement with previous works that study percolation in multilayer networks \cite{2010:buldyrev} and multiplexes \cite{2012:gardenes}.

The interdependent networks formalism developed in this investigation in which the state of each node is the same in every layer can be extended to the study of the spreading of human and animal diseases, the propagation of memes, opinions, rumors, bankruptcies and other situations where agents interact with other agents in several manners but the state of each agent is uniquely determined at every moment. 

The present work was originated in the study of malware spreading, where independent contagion processes take place in every layer concurrently.
Despite of this, the methodology of the effective contagion matrix developed in this work can be applied to other approaches in which such co-ocurrence is limited or absent. The results of such studies will be addressed elsewhere.

We appreciate the useful comments from Jose A.\ Capit\'an. We also want to thank the financial support of MINECO through grants MTM2012-39101 for J.G.\ and FIS2011-22449 (PRODIEVO) for S.C., and of CM through grant S2009/ESP-1691 (MODELICO) for J.G.\ and S.C.


\bibliographystyle{unsrt}

\end{document}